# On The Theoretical Problematic of Arabic Physical Science
# Or
# Why Did Arabic Science Fail To Achieve The Copernican Revolution?

## Hisham Ghassib

Princess Sumaya University for Technology,
Fax: (962 6) 5347295, P.O.Box: 1438 Al-Jubaiha 11941 Jordan.
ghassib@psut.edu.jo

# Abstract

A Hegelianized version of Althusser's concept of problematic is used to investigate the underlying theoretical unity and structure of Arabic physical science (physics, astronomy and chemistry). A contradictory triad ( associated with Platonism, Aristotelianism and Ptolemaism) is identified at the heart of the Arabic project for physical science. The paper focuses on the valiant attempts made by leading Arabic scientists to overcome these contradictions without transcending or tearing apart the prevailing problematic. The following question is then addressed: why was Arabic physical science reformist, rather than revolutionary, unlike Renaissance European physical science? An answer is proposed in terms of the history, nature and decline of Arabic rationalism

# Introduction

My investigations into modern physical theories and their interconnections and mechanisms of development have led me to the idea that the principal mechanism of progress in modern physical theory is a process of unification which dialectically resolves contradictions inherent in the heart of physical theory. Such a process drives physical theory beyond its own structure of premises and meaning; it is a process of transcendence (Ghassib, 1988; Ghassib, 1999).

However, these investigations have also shown that there is always an alternative route of development, which tends to "resolve" contradictions within the existing structure and without transcending it. This route fails to effect major progressive breakthroughs and to open new research avenues, even though it may involve highly sophisticated and clever mathematical ploys and innovative tricks.

I call the first route the revolutionary route, and the second route the conservative route.

The model example which embodies this idea is the events at the end of the nineteenth century which led to the theory of special Relativity. The principal contradiction at the heart of physical theory then was the multi- faceted contradiction between Newtonian Mechanics and Maxwellian Electromagnetism. The conservative route was followed by Lorentz, who "resolved" many aspects of this contradiction, using very sophisticated mathematical techniques, but without transcending the basic structure of Classical Physics epitomized by such notions as the ether, absolute space, absolute time, and absolute mass (Miller, 1986; Einstein et. al., 1980).

Of course, the revolutionary route was followed by Einstein in his 1905 relativity and photon papers. Right from the start, from the very first page of his 1905 relativity paper, Einstein declared his intention to resolve certain "asymmetries" by dispensing with the age-old concept of ether and radically re-examining the concepts of absolute simultaneity and absolute space. This led him to replace these fundamental concepts with the generalized principle of special relativity and the amazing principle of the constancy of the speed of light in vacuo (Einstein, 2005).

We all know where Lorentz's conservative route has led physics (to nowhere) and where Einstein's revolutionary route has led it: to General Relativity, Particle Physics, Modern Cosmology, and the whole of modern physics (Kragh, 2002); Jackson, 1987).

In this paper, I intend to show the efficacy of this idea of development of physical theory in understanding the nature of Arabic physical science, particularly Arabic astronomy, in contrast with Copernican astronomy and the ensuing developments in European science. Focusing on the Arabic critique of Ptolemy and the alternative astronomical models envisaged by various Arabic astronomers, I argue that Arabic scientists tended to follow the conservative route in resolving the contradictions of Greek physical science, whereas Copernicus (and later, Kepler) followed the revolutionary route . In both cases, physics was used as a guide to criticize Ptolemy and develop alternative models. However, whereas Arabic scientists closely followed Aristotelian physics, Copernicus and Kepler openly defied Aristotelian physics and were beginning to feel their way through a new, field, physics (Margolis, 2002).

However, to fully appreciate this characterization, and the nature of innovation in Arabic astronomy, I shall use the Althusserian notion of problematic as a textual tool. However, I find this tool, in its original Althusserian from, too rigid to theorize developmental patterns in scientific theory. I, thus, introduce certain "Hegelianized" modifications to it, which imbue it with noticeable explanatory power.

Finally, I find it necessary to emphasize, right from the start, that the two routes do not grow separately, and in relative isolation from each other. On the contrary, they are inextricably tied to each other. In particular, the revolutionary route is usually unimaginable without the conservative route. Was Einstein imaginable without the conservative Lorentz and the semi – conservative Poincaré? Similarly, was Copernicus imaginable without Al- Zarqali, Al-Bitruji, Ibn Al- Haitham, Al – Urdi, Tusi and Ibn Shatir, amongst others ?

## Louis Althusser (Althusser, 1993; Ferrether, 2006; Elliot, 1994)

Louis Althusser is a French philosopher, who died in 1990. He is usually characterized as a structuralist Marxist philosopher. During the sixties and seventies of the pervious century, he was at the heart of a fierce ideological struggle in French and European left -wing circles. However, with the changed circumstances, and the accompanying retreat of the left worldwide, in the last thirty years, Althusser has suffered an almost total eclipse and been consigned to almost total oblivion. Nevertheless, his influence is still noticeable in cultural studies, particularly after the publication and translation of many hitherto unpublished manuscripts in the last few years (Althusser, 2006). Notwithstanding his ideological and political eclipse, I think that quite a few of his notions could be very effective and useful in understanding historical texts and events, particularly in the sciences. I proclaim Althusser's philosophy as truly pertinent to deepening our understanding of the dialectic of unity and disunity in the history of science. In particular, I deem Althusser's notion of "problematic" to be an essential notion for understanding the unity of Graeco – Arabic physical science vis-a-vis developments in the modern era .

I shall, therefore, explain this notion as a prelude to applying it to Arabic physical science in relation to the Copernican Revolution.

# The Notion of Problematic (Althusser, 1977; Althusser et.al., 1977)

Althusser avers, right at the start of his intellectual career, that the essence of a text does not lie in its object. Nor does it lie in the individual isolated concepts it employs. Rather, it lies in its problematic. The fundamental basis of a text is its problematic, rather than its basic individual concepts. Basically, a problematic is a structured conceptual hierarchy which animates a text and produces its meaning. It is the structural condition of the possibility of meaning of a text. It defines its semantic space. A concept does not acquire its meaning from its logical structure. Nor does it acquire it from a direct relationship to an object outside it. Rather, it does so from its problematic and through it. Even its relationship to its object is established via its problematic.

In fact, Althusser arrived at this notion of problematic by comparing four sets of philosophical texts: Hegel's, Feuerbach's, the Early Marx's and the Mature Marx's. He noticed that, ultimately, the first three sets shared the same problematic, even though they appeared diametrically different. For, even though Feuerbach had inverted Hegel, he had retained the basic underlying Hegelian problematic. Thus, the materialist Feuerbach remained a prisoner of the idealist Hegelian semantic space. Also, even though the Young Marx had been concerned with politics and political economy, whereas Feuerbach had been concerned with religion, theology and speculative philosophy, they both shared the same problematic – or, to be more precise, the Young Marx had borrowed Feuerbach's problematic, and applied it to different objects. Thus, the notion of problematic helps us to detect basic differences and similarities, and to delve deeply into the heart of a text, behind a fascade of illusory appearances. Apparent breaks are recognized for what they truly are – mere variations on a theme. On the other hand, surface appearances and identities turn out to conceal radical breaks and departures. In this respect, the notion of problematic could be used to assess intellectual achievements and their degree of originality.

Althusser also used this important notion to define intellectual revolutions. As long as a text remains tied to the existing problematic, it does not constitute an intellectual revolution, no matter how hard it tries to disguise the problematic with seemingly new concepts, and to invert structures without changing their internal relations. Thus, neither Feuerbach nor the Young Marx constituted an intellectual revolution in philosophy, particularly vis- a-vis Hegel.

On the other hand, the Mature Marx did indeed achieve a radical intellectual revolution, because he succeeded in effecting a so-called epistemological break whereby he broke loose from the Hegelian-Feuerbachian ideological problematic and moved to a new scientific problematic epitomized most conspicuously by Das Kapital. Thus, intellectual revolutions are basically epistemological breaks whereby a thinker jumps from one problematic to an altogether different problematic.

## A Hegelianized Version of Problematics

The moment one tries to apply Althusser's notion of problematic to the history of physical theory, one encounters numerous difficulties related to the obvious rigidity of this notion. Althusser's problematic is indeed too rigid to explain mutations, transitions and movement in physical theory. In fact, it is a closed universe, a monad, with no mechanisms connecting it to other problematics.

Each problematic is almost self-sufficient, coherent and homogeneous. The way out of this impasse for Althusser is the notion of epistemological break– a quantum leap in the dark with no clear mechanism; a blind irrational jump.

The way out to salvage this important notion is to modify it by introducing Hegelian elements in it – in particular, a degree of inhomogeneity and dialectical contradiction. With this modification, a problematic is transformed from a closed static totality into an open dynamic totality that moves forward and mutates under the pressure of its internal contradictions. The state of inhomogeneity in a problematic normally arises from an amalgamation or coalescence of problematics – i.e., from the fact that actual historically constituted problematics are hybrids. This also makes a problematic not indifferent to its object, as Althusser seems to imply. On the contrary, a problematic develops and accentuates its contradictions by interacting with its object, until a point is reached where the resolution of these contradictions demands the transcendence of the problematic. Te be more precise, this resolution transforms the existing problematic into a new higher one.

It is this Hegelianized version of the notion of problematic that I shall now use to explore the nature and significance of Arabic physical Science.

## The Problematic of Greek and Arabic Physical Science

Arabic physical science spans a period of seven continuous centuries (800 A.D. – 1500 A.D.) (Gingerich, 1986; Rashed, 1997). It was indeed a period brimming with innovation and scientific activity .Yet, underlying all this amazing variety of ideas and theories, there was a constant, essentially unchanging, problematic. If one compares a $9^{th}$ century text with a $14^{th}$ century text, one does not fail to notice this constancy in problematic. One, of course, finds it hard to account for this constancy, in view of the many critical spirits that animated Arabic physical science.

How would one characterize the problematic of Arabic Science? Clearly, it was basically a Greek problematic. Arabic physical Science was principally a creative continuation of Greek physical Science. In particular, it was an amalgam of three "pure" problematics– Plato's, Aristotle's and Ptolemy's. They are three inter–related, but distinct, pure problematics. Plato's problematic, as revealed in the Timaeus (Plato, 1978), revolves around the idea that the Universe is a unique, self–contained, self-sufficient and rational being, endowed with perfect traits and features, such as a spherical shape and components revolving around its centre with uniform circular motion. The latter traits became a cornerstone of physical science, astronomy and cosmology in Antiquity. It was deeply incorporated in the Aristotelian problematic (Sarton, 1966). The latter was based on the idea that the Universe is a semi-material, finite, inhomogeneous, mechanical system with a well – defined physical centre, that acts as a gathering and attractive place for heavy elements (earth and water). This spherical system knows no outside and no vacuum. It consists of two distinct and qualitatively different realms: the terrestrial and the celestial. The former is characterized by change, corruption, birth, death, straight-line motion and a combination of four basic elements (earth, water, air and fire), and is described by Aristotelian Physics. The latter is characterized by sphericity, uniform circular motion, and an eternally unchanging substance or element (ether). Space is a mere attribute of matter, and time is endless. The planets and stars are carried by revolving spheres made of transparent ether. There is a prime unmoved mover that envelopes the whole Universe and imparts motion to the various concentric spheres (Aristotle, 1978).

Ptolemy later developed a mathematical problematic, based on work done previously by Appollonius, Hipparchus and the Babylonians (Barbour, 2001). He employed such geometrodynamic concepts as the eccentric, deferent, epicycle and equant. In fact, as an astronomer seeking to describe and "explain" data and measurements, he was truly revolutionary in light of later developments, especially Kepler's work. However, as a physicist, he was truly conservative, true to the Aristotelian problematic. He hoisted his revolutionary mathematical problematic onto the Aristotelian problematic, creating an explosive, contradictory amalgam that would stamp the dynamic of physical science for the following millennium and a half.

## The Dilemmas and Dynamics of Arabic  Physical Science

The problematic Arabic physical scientists inherited from the Greeks was an inhomogeneous, contradictory problematic. The principal contradiction was the multifaceted contradiction between the Ptolemaic mathematical apparatus and Aristotle's physics. Arabic astronomers felt dissatisfied with Ptolemaic methods on account of this contradiction and some observational flaws they discovered in Ptolemy's system.

Basically, there were three main currents or traditions in Arabic astronomy: the observational (zij), the Shukuk tradition enunciated by Al-Hassan ibn Al – Haitharn,  and the model- building tradition ( Bitruji, Urdi,

Tusi, Ibn Shatir). They all moved and produced ideas within the confines of the inherited problematic (Saliba, 1994). In spite of their highly critical attitudes and creative talents, none ventured to move forward beyond it. The models the model-builders constructed fitted better in the edifice of the main Aristotelian problematic than Ptolemy's models. All Arabic astronomers, including those belonging to the zij tradition, were acutely aware of the contradictory status of their problematic, and they tended to use their great mathematical skills to resolve those contradictions, guided by Aristotelian principles .

In a sense, Arabic astronomers developed the conservative elements in the ancient problematic at the expense of the revolutionary elements; Aristotle at the expense of Ptolemy. They evidently opted for the conservative route of development. Their efforts were directed at taming the rebellious elements in Ptolemy. Ibn Al-Haitham's Shukuk found their consummation in the Maragha School; in Urdi's, Tusi's and Ibn Shatir's models and mathematical innovations. Of course, they were unable to dispense completely with Ptolemy's mathematical methods. So, they resorted to a selection process according to taste within the confines of Aristotelian principles. They all agreed on the necessity of dispensing with the most revolutionary element in Ptolemy– namely, the equant. As we know, Tusi was able to replace it with his famous "couple". Ibn Shatir built models that were free from equants and eccentrics, but that contained deferents and epicycles, and he had to modify slightly Aristotle's theory of the ether to justify that. It was as though Arabic astronomers used all their mathematical ingenuity and genius to save the Greek problematic from its own contradictions. Their ideal was to create purely Aristotelian planetary models that accorded well with the observations. The purest Aristotelian system was built by Al-Bitruji in Andalus, but turned out to be inaccurate, compared with Ptolemy's (Al-Bitruji, 1971). Thus, Arabic astronomers had no choice but to retain some Ptolemaic elements.

If we compare ancient astronomy to physics at the end of the 19[th] century, we could characterize the situation as follows: Ptolemy was the Planck of ancient astronomy, Arabic astronomers were its Lorentz, whilst Copernicus, Galileo and Kepler were its Einstein.

The pivotal questioning that would explode the ancient problematic along the revolutionary route was, of course, that related to the Earth–centered hypothesis. Copernicus readily adopted the solutions that had been offered by Arabic astronomers, because they satisfied his Platonic aesthetic taste. In this, he was as conservative as Arabic scientists. However, he differed from them in being a Platonist, rather than an Aristotelian (Kuhn, 1981; Koyre, 1973). That must have facilitated his heliocentric revolution. It was indeed a revolution, because it challenged the whole Aristotelian edifice, and pointed towards a new, field, physics– modern physics. Thus, both Arabic astronomers and their European counterparts used physics to guide their astronomical practice. However, whereas the former tended to stick to Aristotelian physics in this endeavour, the latter tended to challenge it and use rudiments and intimations of a new, field, physics to guide their model-building. In particular, Kepler used a field physical model, inspired by Gilbert's magnetic force model, to investigate the way the sun influenced the motion of the planets, and that played a crucial role in arriving at his laws of planetary motion. In fact, it enabled him to discover the revolutionary content of the Ptolemaic equant– the fact that it was a first approximation of an elliptical orbit (Stephenson, 1987). Thus, unlike Arabic astronomers, who had rejected the equant and retained the epicycle, he opted for the former and rejected the latter. He proved to be the ultimate embodiment of the revolutionary route.

The preceding analysis inevitably raises the following questions: Why did Arabic astronomers follow the conservative Lorentzian route, despite their acute critical acumen and brilliant creative mathematical skills? Why did the Copernican Revolution have to await a Renaissance Central European – or, more precisely, two such astronomers (Copernicus and Kepler)? Why did Arabic astronomers stick so stubbornly to Aristotelian physics, and felt so inimical towards the revolutionary elements in Ptolemy?

## The Rise and Demise of Arabic Rationalism

It is our contention that Arabic theoretical astronomy was closely related to Arabic rationalism, which means that the fate of the former was organically tied to the fate of the latter. My thesis here is that Arabic rationalism, which had been adopted by the early Abbasid state, particularly during the reigns of Al-Mansour and Al-Ma'moun, was soon to come under fire from orthodox religious quarters (Al-Jabiri, 1985). The ascendancy of this orthodoxy and the alliance it forged with the militarist feudalist Islamic state weakened Arabic rationalism, and placed it under constant siege. Eventually it was liquidated completely by this alliance. Its last bastion was Arabic astronomy. The rationalists felt constantly threatened by that inimical force. This explains why they stubbornly stuck to Aristotelian rationalism, and were reluctant to transcend it. The fact that the battle for rationalism was lost in Arabic civilization accounts for this stubborn adherence to Aristotelian physics as opposed to Ptolemaic innovations. Because Arabic rationalists were under siege in their last bastion, their chief task was to defend Aristotelian rationalism, and not to critique it. That also offers a partial explanation of why Arabic astronomy reached its zenith in the age of decline.

Arabic rationalism declined and was eventually liquidated for various socio-cultural reasons, one of which was its failure to forge a compromise with religious dogma and to carve a niche for itself in the Islamic religious enterprise. It was a double failure, whereby each enterprise failed to accommodate itself to the other. In Europe, the exact opposite occurred. In spite of the initial hostile response of the Church to Aristotelian and Averroist rationalism (Grant, 1971; Grant, 1974), eventually a historic compromise was reached between the two enterprises, and Aristotelianism was absorbed by Church dogma, thanks chiefly to the brilliant efforts of Thomas Aquinas. The fact that Aristotelian rationalism became part of the official dogma made it possible for later generations to critique it in the name of a higher rationalism. The consolidation of Aristotelian rationalism in European civilization was a precondition for its thorough critique in the name of a higher rationalism. This difference in attitude is best illustrated by contrasting Kepler's attitude to Ptolemy's equant with the late Arabic astronomers' attitude to it. Arabic astronomers were vehement in rejecting this Ptolemaic device, because it contradicted Aristotelian physics, which was part of Aristotelian rationalism. Their strict normative adherence to this rationalism drove them to view the equant as an irrational element in Ptolemaic astronomy. Of course, Copernicus was to follow suit. On the other hand, Kepler was to reject the Ptolemaic epicycle and rehabilitate the Ptolemaic equant in the name of a new, burgeoning, physics and rationality. The triumph of Aristotelian rationalism in medieval Europe was a prelude to its later demise and its replacement by a new rationalism, what I call scientific rationalism. On the other hand, its defeat in Arabic medieval civilization was an impediment to transcending it towards a new science and a new rationalism. It was a prelude to the eclipse of rationalism as such in Arabic civilization.

The besieged status of Aristotelian rationalism drove Arabic astronomers to seek solutions to the problems of theoretical Arabic astronomy within the confines of Aristotelian physics and cosmology, just as Lorentz later did with respect to classical electrodynamics (Pais, 2008). The consolidation of Aristotelian rationalism in Arabic civilization would have been a necessary precondition for its transcendence towards modern scientific rationalism— i.e., for the Copernicus-Kepler astronomical revolution to occur in the Arabic middle ages.

## Metaphysical Rationalism Versus Scientific Rationalism

To understand the significance of this contrast and the associated intellectual processes, we need to understand the distinguishing features of both antique rationalism—i.e., the rationalism that prevailed in both Greek and Arabic antiquity—and modern rationalism, which has prevailed since the scientific revolution (1543-1687).

I call antique rationalism metaphysical rationalism. In Arabic civilization, this rationalism existed in quite a few forms: Aristotelian or Peripatetic, Platonic, Neo-Platonic (Plotinus) and Kalamic ( both Mu'tazilite and Ash'arite). These rationalist forms were combined with various forms of irrational currents, such as Pythagoreanism, mysticism, astrology, alchemy, magic, oriental creeds, gnosticism and hermeticism (Al-Jabiri, 1986). This rationalism reached its apex in Aristotelian rationalism, and, in particular, in Ibn Rushd (Al-Jabiri, 2001). In Atistotelian rationalism, there were two sources of certainty: pure reason and the raw senses. A set of metaphysical principles were derived from pure reason, and considered solid bases of certain knowledge, on the one hand, and the world of sense was taken uncritically for granted, on the other. The essence of Aristotelian physics lay in emphasizing nature related metaphysical principles, and, then, saving the phenomena with them. Attempts were made to account for the raw phenomena of the world of the senses in terms of these metaphysical principles via formal logic and the syllogism. The latter was the bridge between metaphysics and the senses. The world of metaphysical principles and the world of the senses were related via the syllogism. That was the essence of Aristotelian methodology and metaphysical rationalism.

A new rationalism was to arise during the 17$^{th}$ century—scientific rationalism. Whereas Arabic astronomers and physicists failed to transcend metaphysical rationalism to scientific rationalism, conditions were ripe for 17$^{th}$ century European scientists to do that. Basing our conclusions on a thorough study of the 17$^{th}$ century scientific revolution, we outline here some of the distinguishing features of scientific rationalism (Ghassib, August, 2010):

1- The relationship between philosophy and science was inverted. Instead of science being appended to metaphysics, generally speaking, metaphysics and philosophy became appended to science. Instead of philosophy guiding science, from then on, science would guide philosophy.

2- Whereas antique science focused on the actual as revealed by the senses, and in the light of the metaphysical abstract, modern science would focus on the potential, the possible, the mathematically amenable abstract. Whereas the former tried to find out how first (metaphysical) principles produced raw phenomena, modern science would try to discover the mathematical principles followed by the potential and possible. The latter would then dialectically fuse these principles together and universalize them to arrive at deeper material causal principles. In short, mathematical deduction and dialectical synthesis have replaced the syllogism, on the one hand, and material causes have replaced first principles and metaphysical causes, on the other

3- In antique science, philosophy was endowed with the function of producing knowledge. It was a necessary tool of knowledge production. In modern knowledge production, it has lost this function, and the latter has been replaced by the function of transcendentally grounding scientific practice, and other practices as well.

4- Science has become autonomous. In particular, it has gained its independence of theology, philosophy, the crafts and political authority. Scientific reason has become an authority unto itself.

5- Notwithstanding quantum mechanics, modern science presupposes the ideas of the materiality and infinitude of the world. It also presupposes that causes are internal to the Universe, and a new causality based on material interactions between material components.

6- In modern science, the world of the senses is no more a source of certainty. It is not considered entirely objective, but partly subjectively constituted. The senses are basically defective measurement instruments. Knowledge needs more accurate and precise measuring instruments. These are furnished, not by the senses, but by scientific reason, which is, thus, the eye of truth and our probe of reality. Modern science presupposes the existence of primary and secondary qualities and properties—i.e., it presupposes the dichotomy of appearance and reality.

7- Modern knowledge is mathematically structured. This means that it is axiomatically based. It presupposes a set of axioms. However, scientific axioms are not absolute, but relative and conditional. They are testable and studied presuppositions. They are constantly under scrutiny and constantly tested via their deductive consequences. They are always provisional. In metaphysical rationalism, on the other hand, axioms have an absolute and sacred character.

8- The mathematical experimental essence of scientific rationalism imposes a set of ethical norms on scientific practice, which is alien to metaphysical rationalism.

9- Modern science and rationalism are essentially absolutely critical, dialectical, unifying, materialist and causal in the materialist sense. Metaphysical rationalism tended to be dogmatic, formally syllogistic, fragmenting, idealist and causal in the theological sense.

10- The animating principle of scientific practice and rationalism is scientific reason. Science has no reference point save scientific reason. The animating principle of metaphysical rationalism was transcendental reasoning.

## Conclusion

We have hegelianized Althusser's concept of problematic, and applied it to Arabic astronomy, to uncover the underlying unity of that science. We have shown that Arabic astronomers chose to follow the conservative route of development a la Lorentz, unlike Kepler, who decided later to follow the revolutionary route a la Einstein. We, then, offered an explanation of why that occurred, in terms of a contrast we have detailed between metaphysical rationalism and scientific rationalism.